\documentclass[aps,pre,preprint,superscriptaddress,amsmath,amssymb,nofootinbib]{revtex4}
\usepackage{amsmath}
\usepackage{graphicx}
\usepackage{amssymb}
\usepackage{dcolumn}
\usepackage{bm}
\usepackage{color}
\usepackage{float}
\usepackage{upgreek}
\usepackage[final]{hyperref} 
\hypersetup{
	colorlinks=true,       
	linkcolor=blue,        
	citecolor=blue,        
	filecolor=magenta,     
	urlcolor=blue         
}

\usepackage{upgreek}
\usepackage[normalem]{ulem}

\DeclareMathOperator\erf{erf}

\makeatletter

\begin{document}
\title{Confined diffusion in a random Lorentz gas environment}
	
\author{Narender Khatri}\thanks{narenderkhatri8@iitkgp.ac.in}
\affiliation{Department of Physics, Indian Institute of Technology Kharagpur, Kharagpur - 721302, India}

\author{P.S. Burada}\thanks{Corresponding author: psburada@phy.iitkgp.ac.in}
\affiliation{Department of Physics, Indian Institute of Technology Kharagpur, Kharagpur - 721302, India}
\affiliation{Center for theoretical studies, Indian Institute of Technology Kharagpur, Kharagpur - 721302, India}

\date{\today}
	
\begin{abstract} 
We study the diffusive behavior of biased Brownian particles in a two dimensional confined geometry filled with the freezing obstacles.
The transport properties of these particles are investigated for various values of the obstacles density $\eta$ and the scaling parameter $f$, which is the ratio of work done to the particles to available thermal energy.
We show that, when the thermal fluctuations dominate over the external force, i.e., small $f$ regime, particles get trapped in the given environment 
when the system percolates at the critical obstacles density $\eta_c \approx 1.2$.
However, as $f$ increases, we observe that particles trapping occurs prior to $\eta_c$.
In particular, we find a relation between $\eta$ and $f$ 
which provides an estimate of the minimum $\eta$ up to a critical scaling parameter $f_c$ beyond which the Fick-Jacobs description is invalid.
Prominent transport features like nonmonotonic behavior of the nonlinear mobility, anomalous diffusion, and greatly enhanced effective diffusion coefficient are explained for various strengths of $f$ and $\eta$. 
Also, it is interesting to observe that particles exhibit different kinds of diffusive behaviors, i.e., subdiffusion, normal diffusion, and superdiffusion.
These findings, which are genuine to the confined and random Lorentz gas environment, can be useful to understand the transport of small particles or molecules in systems such as molecular sieves and porous media which have a complex heterogeneous environment of the freezing obstacles. 
\end{abstract}

\maketitle

\section{Introduction}

Diffusion of small particles or molecules in a crowded environment is ubiquitous in several physical, chemical, and biological processes \cite{Lowen_RMP,Dix}.
Often, these particles encounter a heterogeneous environment 
\cite{Golding, Drescher,Moeendarbary,Lowen,Ghosh_pre} formed by obstacles distributed in an irregular fashion. This may control the diffusive behavior of the particles. 
Examples are, particle diffusion in biological environments containing lipids and proteins \cite{Dix,Ellis}, diffusion in living cells crowded with cytoplasmic and nuclear environments \cite{Minton}, diffusion in porous soil columns \cite{Cortis}, etc.
From the theoretical perspective, the complex heterogeneous environment is modeled as a random Lorentz gas \cite{Weijland, Hofling,Zeitz}, where the freezing obstacles are randomly distributed with a given area fraction. 
The properties of this random Lorentz gas depend entirely on the obstacles density.
For example, at the higher obstacles density, percolating clusters \cite{Mertens_pre} will form, which effectively control the diffusive behavior of the particles.
Note that when the system percolates at the critical obstacles density, the clusters can even arrest or trap the particles \cite{Zeitz,Morin_pre}. 

On the other hand, when the particles diffuse in a confined geometry, their motion is highly controlled by the structure of the geometry \cite{Karger,Schmid,Rubi,Burada_cpc,Borromeo,Khatri,Talbot}, e.g., 
ion channels \cite{Hille}, 
zeolites \cite{Barrer}, 
microfluidic devices \cite{Han}, 
ratchets \cite{Matthias,Kettner,Marchesoni_rev,Burada_prl2,Li_pre} 
and artificial channels \cite{Reza}.
The irregular shape of the structure gives rise to entropic barriers which play a prominent role in the diffusive behavior of the particles \cite{Jacobs,Zwanzig,Burada_prl,Burada_phd,Marten,Pineda,Ai_jcp,Das_jcp,Hanggi_PNAS,Burada_pre}. In these confined geometries, the transport characteristics are controlled by the effective free energy, which is a function of applied bias and the entropic potential \cite{Burada_cpc, Khatri, Burada_prl, Burada_phd,Marten}. 
When the shape of the geometry is periodic and regular, 
to analytically calculate the transport characteristics of the non-interacting particles,
one can use the Fick-Jacobs theory \cite{Jacobs, Zwanzig}, which assumes a faster equilibration of the diffusing particles in the transversal direction of the channel compared to its longitudinal direction. 
The prominent transport features reported in these structures include a decrease in average particle velocity upon increasing the noise strength and exhibiting an enhanced effective diffusion coefficient in highly confined geometries 
\cite{Burada_prl, Burada_phd,Marten,Burada_pre}.
However, quite often, particles encounter a crowded environment while passing through the confined structures \cite{Minton, Benichou_Condens,Conrad} such as biological cells, 
microfluidic channels, blood vessels, and porous media.   
Due to the combined effect of density of obstacles and structure of the confined geometry, 
the transport properties of the particles may exhibit interesting behaviors.

In this article, we study the diffusive behavior of point size particles, 
moving in a two dimensional confined channel filled with the freezing obstacles. 
These particles are subjected to a constant external bias along the channel direction. 
Here, we consider the steric interaction between the diffusing particles and the freezing obstacles. We aim to find the transport characteristics of the Brownian particles, i.e., the nonlinear mobility and the effective diffusion coefficient, in the aforementioned conditions.

Rest of this article is organized as follows. In section -\ref{Model}, we introduce our model for the biased Brownian particles in a two dimensional confined channel with a random Lorentz gas environment. The impacts of such a confined and crowded heterogeneous environment on the nonlinear mobility and the effective diffusion of the particles are presented in section -\ref{Mobility} and section -\ref{Diffusion}, respectively. Finally, we present our main conclusions in  section -\ref{Conclusions}.

\section{Model}\label{Model}

\begin{figure}[thb]
\centering
\includegraphics[scale = 1.4]{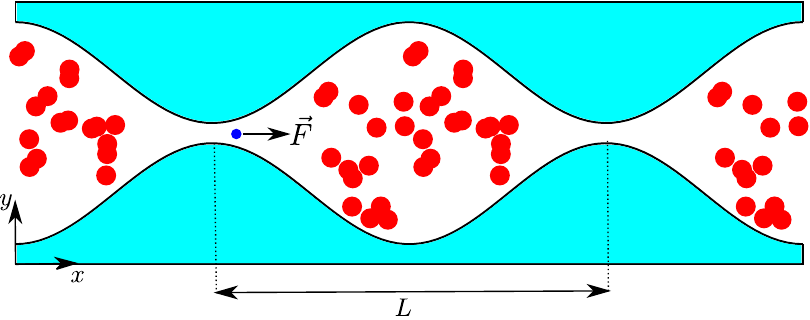}
\caption{(Color online) Schematic illustration of a two dimensional symmetric channel in which there is a heterogeneous environment of freezing obstacles modeled by a random Lorentz gas, with the periodicity $L$ confining the motion of a Brownian particle which is subjected to a constant force $\vec{F}$ along the $x-$direction. 
The reflecting channel boundaries assure the confinement of a particle inside the channel.}
\label{fig:chw}
\end{figure}

Consider an overdamped Brownian particle suspended in a two dimensional symmetric channel consisting of a heat bath of friction coefficient $\gamma$ at the temperature $T$. 
Also, a heterogeneous environment of freezing obstacles (spatial Poisson point process) \cite{Stauffer}, 
modeled by a random Lorentz gas, is present inside the channel as illustrated in Figure~\ref{fig:chw}. 
The shape of the two dimensional symmetric and spatially periodic channel is described by its half-width (see Fig.~\ref{fig:chw}) 
$\omega(x)= a\sin\left(2\pi x/L \right)  + b,$
where $L$ corresponds to the periodicity of the channel, and the parameters $a$ and $b$ control the slope and channel width at the bottleneck, respectively. 
Here, we choose $a = 1/2\pi$, $b = 1.02/2\pi$, and $L = 1$.
 
The particle is driven by a constant external force $\vec{F}$ along the direction of the channel and the interaction force $\vec{F}_\mathrm{int}$ due to the obstacles. 
In the overdamped regime \cite{Purcell}, the equation of motion of the particle is given by the 2D Langevin equation,
\begin{equation}
\gamma \frac{d \vec{r}}{d t} = \vec{F} + \vec{F}_\mathrm{int} +  \sqrt{\gamma k_B T} \vec{\xi}(t),
\label{eq:Langevin1}  
\end{equation}
where $\vec{r}$ is the position of the particle in two dimensions and $k_B$ is the Boltzmann constant. The thermal fluctuations due to the coupling of the particle with the surrounding heat bath are modeled by a zero-mean Gaussian white noise $\vec{\xi}(t)$, with the autocorrelation function $\langle \xi_i (t)  \xi_j (t') \rangle = 2 \delta_{ij} \delta(t-t')$ for $i, j = x, y$.    

As the channel geometry is periodic along the longitudinal direction, we use the periodic boundary conditions. If each obstacle has the same radius $R_0$, then the obstacles occupy an area fraction $\eta = n \pi R_{0}^{2}/A$, where $A = 2bL$ is the area of a single cell of the channel and $n$ denotes the total number of obstacles in the cell. 
However, the obstacles can fully overlap each other, thus the actual area fraction is given by \cite{Torquato} 
\begin{equation}\label{areaf}
\phi = 1 - e^{-\eta}. 
\end{equation}

The interaction force on a particle $i$ due to the freezing obstacles is assumed to be of the linear spring form \cite{Ai_H}, which reads
\begin{equation}\label{Harmonic}
\vec{F}_\mathrm{int} =  k_\mathrm{s}\sum_{j=1}^{n}(R_0 - r_{ij})\hat{r}_{ij}, \quad \text{for} \quad r_{ij} < R_0, 
\end{equation}
where the sum is taken over all the obstacles within the cell in which the particle is present at the given instant of time, $k_\mathrm{s}$ denotes the spring constant, and $r_{ij}$  denotes the center to center distance between the particle $i$ and the obstacle $j$. For $r_{ij} < R_0$, the obstacle strongly repels the particle, however, for $r_{ij} \geq R_0$, there is no interaction between the obstacle and the particle. In order to mimic hard-core pure volume exclusion, we use a large value of $k_\mathrm{s}$, ensuring that the particle does not overlap with the obstacles.

For the sake of a dimensionless description, we henceforth scale all lengths by the periodicity of the channel $L$ and time by  $\tau = \gamma L^2/(k_B T)$. 
In dimensionless variables, the 2D Langevin equation (Eq. \ref{eq:Langevin1}) reads
\begin{equation}\label{Langevin2}
\frac{d \vec{r}}{d t} = \vec{f} +  \vec{f}_\mathrm{int} + \vec{\xi}(t),
\end{equation}
where $\vec{f} = f \hat{x}$ ($f = F L/k_B T$) denotes the dimensionless external force, which is the ratio of work done on the particle due to the external force and the available thermal energy, and the dimensionless interaction force becomes $\vec{f}_\mathrm{int} = \vec{F}_\mathrm{int} L/k_B T$. 
For our simulations, we consider the Brownian particles of point size with random initial conditions and overlapping with the obstacles is taken care by the interaction force $\vec{f}_\mathrm{int}$. The Langevin equation (Eq. \ref{Langevin2}) is solved by using the standard stochastic Euler algorithm over $5 \times 10^2$ trajectories with a time step $10^{-8}$. The reflecting channel boundaries assure the confinement of particles inside the channel. For the simulations, we have chosen the dimensionless spring constant $k$ 
($k=k_\mathrm{s} L/k_B T$) $= 5 \times 10^6$ and the dimensionless radius of the obstacle $R_0 = 0.02$.

\section{Nonlinear Mobility}\label{Mobility}
 
In the absence of a random Lorentz gas environment, i.e., $\eta = 0$, the nonlinear mobility of biased Brownian particles in a symmetric confined environment has been studied earlier both analytically and numerically by H\"anggi and co-workers \cite{Burada_prl,Burada_phd,Marten,Burada_pre}.
It has been shown that under the assumption of a fast equilibration in the transverse direction of the channel, the 2D Smoluchowski equation can be reduced to an effective 1D equation (the Fick-Jacobs equation), reading in the dimensionless form \cite{Burada_prl,Burada_pre}
\begin{equation}
\frac{\partial P(x, t)}{\partial t} = \frac{\partial}{\partial x} D(x) \left\{\frac{ \partial P(x, t)}{\partial x} + A'(x) P(x, t)\right\}, 
\end{equation}
where $P(x, t)$ denotes the reduced probability density, $D(x) = 1/\left( 1 +  \omega' (x)^2 \right)^{1/3}$ is the position dependent diffusion coefficient for a 2D system \cite{DR_pre}, the dimensionless free energy is given by $A(x) =  -f x - \ln (2 \omega (x))$, and $2\omega (x)$ is the local width of the two dimensional channel.
Note that, here, the free energy assumes the form of a periodic tilted potential whose barrier height is a function of the temperature \cite{Burada_prl}. Using the mean first passage time (MFPT) approach \cite{Burada_prl, Burada_phd}, the nonlinear mobility can be obtained as 
\begin{equation}\label{mob}
\mu := \frac{\langle \dot{x} \rangle }{f} 
= \frac{1 - e^{-f}}{f \displaystyle{\int_{0}^{1} I(z) \, dz}}, 
\end{equation}
where the integral $I(z) = e^{A(z)}/D(z) \int_{z-1}^{z}  e^{-A(y)} \, dy $.
The same quantity is calculated using the numerical simulations for the 2D channel as 
\begin{equation}\label{nonlinear mobility}
\mu := \lim_{t\to\infty} \frac{\langle x(t) \rangle}{ t \, f}.
\end{equation}
 
 \begin{figure}[t]
\includegraphics[width = 0.6\linewidth]{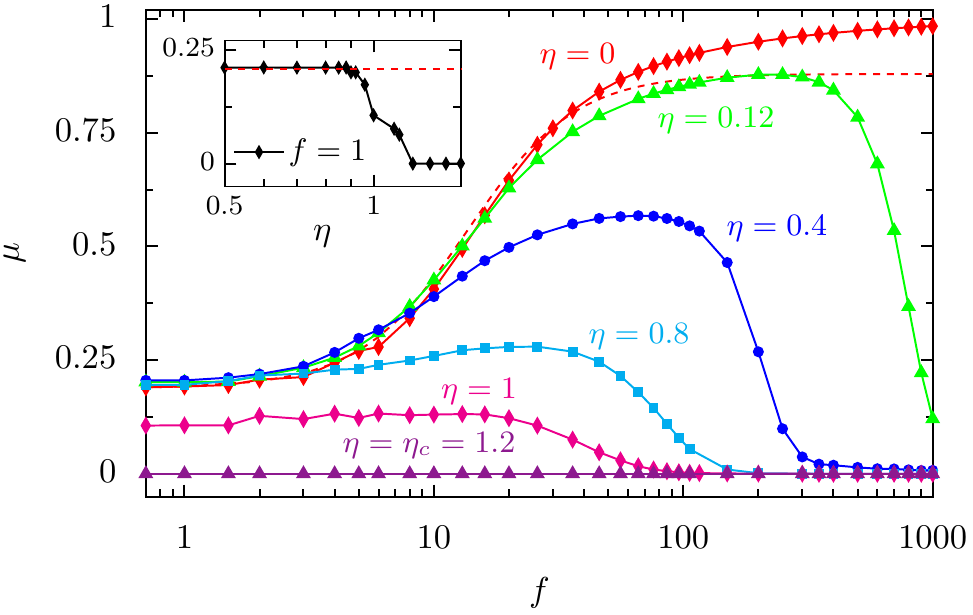}
\caption{(Color online) The nonlinear mobility $\mu$ versus the scaling parameter $f$ for various values of the obstacles density $\eta$. The dashed line corresponds to the analytical findings for $\eta = 0$. The inset depicts the dependence of $\mu$ on $\eta$ for $f = 1$. The other set parameters are $a = 1/2\pi$, $b = 1.02/2\pi$, and $R_0 = 0.02$.}
\label{fig:Mobility}
 \end{figure}

\begin{figure}[t]
\includegraphics[width=0.8\linewidth]{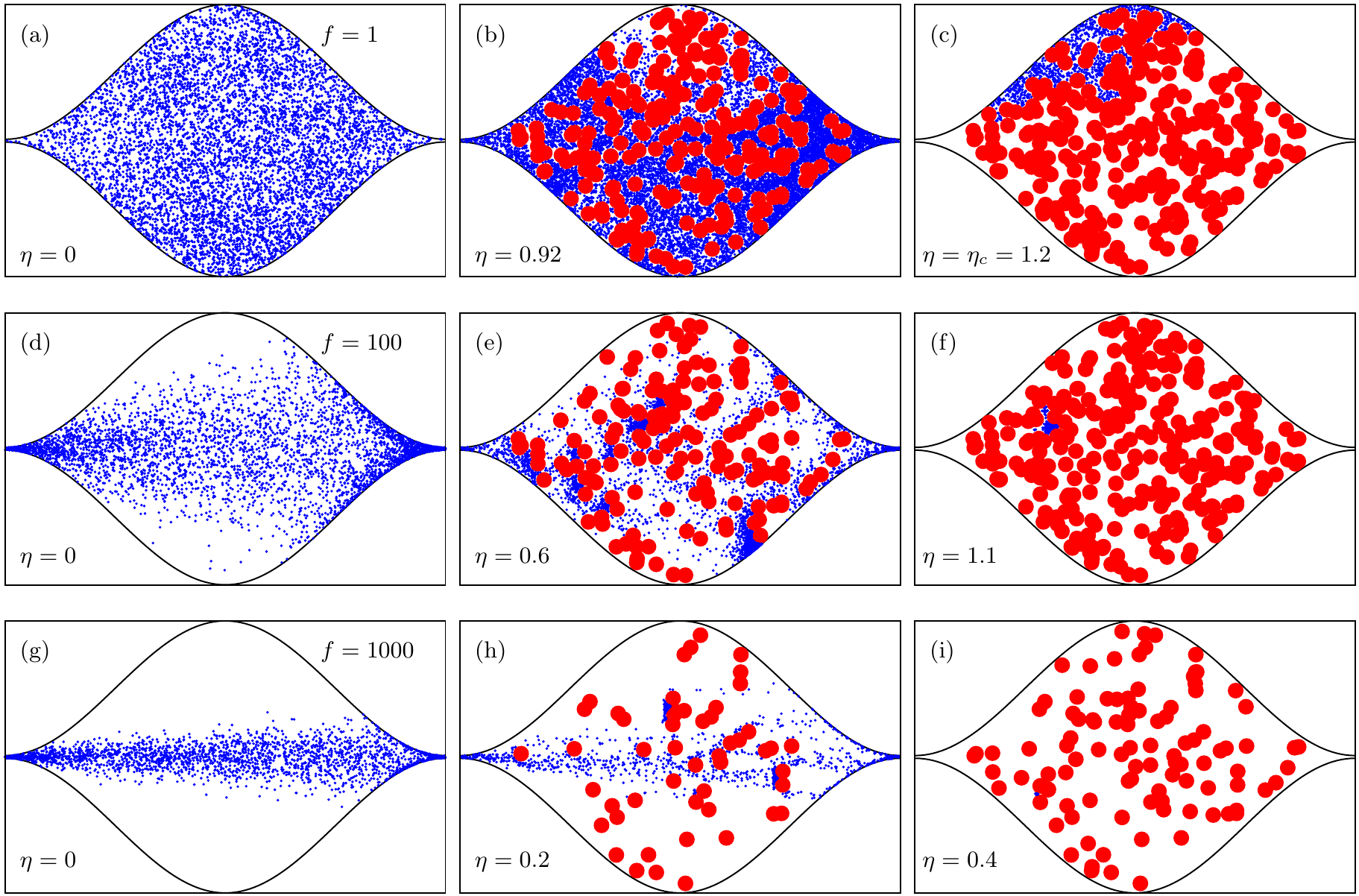}
\caption{(Color online) Steady state evolution of a Brownian particle, 
within a finite time window, for various values of the obstacles density $\eta$. 
Here, the position of the particle is mapped into a single cell of the 2D channel filled with the heterogeneous distribution of obstacles. 
Top panel (a)-(c) for $f = 1$, middle panel (d)-(f) for $f = 100$, and bottom one (g)-(i) for $f = 1000$. The other set parameters are $a = 1/2\pi$, $b = 1.02/2\pi$, and $R_0 = 0.02$.}
\label{fig:Distibution}
\end{figure}

Figure~\ref{fig:Mobility} depicts the nonlinear mobility $\mu$ as a function of the scaling parameter $f$ for various values of the obstacles density $\eta$.
The nonlinear mobility is greatly influenced by the obstacles density. 
For lower values of $\eta \, (< 0.12)$, $\mu$ monotonically increases with $f$ and attains the bulk value in the limit $f \rightarrow \infty$. 
As expected, the numerical simulations results are in good agreement with the analytical findings when $\eta \rightarrow 0$ with deviations occurring at higher values of the scaling parameter $f$. It is due to the failure of the thermal equilibration assumption in this limit.
Interestingly, on further increasing $\eta$, $\mu$ exhibits a nonmonotonic behavior. It is because, for higher $f$ values, the freezing obstacles slow down the motion of particles and the mobility becomes zero.
Surprisingly, in the lower $f$ limit, the nonlinear mobility is unaltered and agrees well with the analytical findings up to the obstacles density $\eta \approx 0.92$ (see inset of Fig.~\ref{fig:Mobility}). 
This indicates that, in this limit, the obstacles do not have much impact on the diffusion of particles and the thermal equilibration assumption is still valid (see Fig.~\ref{fig:Distibution}(b)). 
Here, the thermal fluctuations dominate, and the particles can move freely in the 2D channel. However, at the critical obstacles density $\eta_c \approx 1.2$, obstacles form percolating clusters that span the cell of the 2D channel. As expected, the particles get trapped or localized in these clusters, and the nonlinear mobility becomes zero (see inset of Fig.~\ref{fig:Mobility}). 
Note that in a two dimensional square box the critical obstacles density for the percolation of disks is $\eta_c \approx 1.128$ \cite{Mertens_pre}, while for our system it is obtained as $\eta_c \approx 1.2$. 
This indicates that $\eta_c$ depends on the shape of the channel structure. 

In general, according to the ergodic hypothesis \cite{Berkovitz}, the steady state evolution of a Brownian particle within a time window is equivalent to the steady state  distribution of the particles.
Figure~\ref{fig:Distibution} depicts the steady state behavior of a single particle in a finite time window.
Note that since the channel is periodic and the distribution of obstacles in a given cell is heterogeneous, we have mapped the position of the particle into a single cell.
When $f$ value is small, the particle can explore uniformly in the transversal direction of the cell, and this indeed satisfies the thermal equilibration assumption \cite{Burada_phd,Burada_pre} up to $\eta \approx 0.92$ (see (a)-(b) in Fig.~\ref{fig:Distibution}).   
This reflects the fact that, in the lower $f$ limit where the thermal fluctuations dominate, the obstacles do not influence the particle diffusion. Whereas, by further increasing $\eta$, the thermal equilibration assumption is no longer valid because the percolating clusters hinder the diffusion of the particle in both the longitudinal and transversal directions. 
Note that when the system percolates at the critical obstacles density $\eta_c \approx 1.2$, the particle gets trapped in the given environment (see Fig.~\ref{fig:Distibution}(c)).
However, for the moderate and higher $f$ values, i.e., when the thermal fluctuations are less dominant, 
the particle gets trapped in the given environment well before the critical obstacles density $\eta_c$ (see (f) and (i) in Fig.~\ref{fig:Distibution}). 
For the moderate $f$ value the particle gets trapped at $\eta \approx 1.1$, and 
for the higher $f$ value this occurs at $\eta \approx 0.4$.
As reported earlier \cite{Burada_pre}, in $\eta \rightarrow 0$ limit, for the moderate and higher $f$ values, the Brownian particle evolution tends to focus at the middle and exit of the channel evidencing the failure of the thermal equilibration assumption (see (d) and (g) in Fig.~\ref{fig:Distibution}). 
This is because the influence of the external force, which tries to drag the particle in its direction, is much more effective than the thermal noise present in the system.

\begin{figure}[t]
\includegraphics[width=0.6\linewidth]{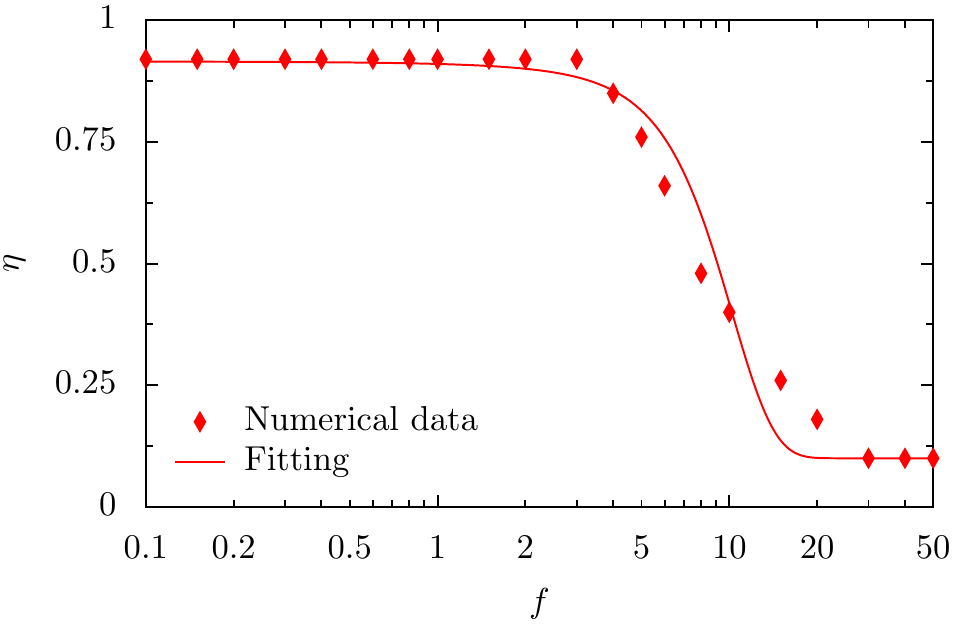}
\caption{(Color online) The $\eta$ versus $f$ relation which provides an estimate of the minimum $\eta$ up to a critical scaling parameter $f_c \sim 51$ beyond which the Fick-Jacobs description is not valid. 
Symbols represent numerical data, and the solid line is an empirical relation $\eta(f)= c_1 - c_2\, \erf((f - \lambda)/\Sigma)$, where 
the parameters are $c_1 = 0.51, c_2 = 0.41, \lambda = 9$, and $\Sigma = 5$.
The other set parameters are $a = 1/2\pi$, $b = 1.02/2\pi$, and $R_0 = 0.02$.}
\label{fig:FJ_Validation}
\end{figure}

In the absence of a random Lorentz gas environment, i.e., $\eta = 0$, Burada \textit{et al.} \cite{Burada_pre} analytically obtained a critical scaling parameter, which is given by $f_c \sim L^2 [1 - \langle \omega '(x)^2 \rangle]/(2 \langle \omega(x)^2 \rangle)$, where $\langle \cdots \rangle$ denotes the average over the period $L$ of the channel. This critical scaling parameter $f_c$ provides an estimate of the minimum forcing beyond which the Fick-Jacobs description is expected to fail in providing an accurate description of the dynamics of the system. For the considered channel structure, we find that $f_c \sim 51$, up to which the numerical simulations results are in good agreement with the analytical findings when $\eta \rightarrow 0$ (see Fig.~\ref{fig:Mobility}). 
Surprisingly, as mentioned earlier, we observe that there exist specific values of $\eta$ up to $f_c$ below which the thermal equilibration assumption is naturally satisfied. In other words, there exists a $\eta$ versus $f$ boundary below which the numerical results agree well with the analytical finding for $\eta = 0$ 
(see Fig.~\ref{fig:FJ_Validation}). 
Unfortunately, in the presence of a heterogeneous environment inside the channel formed by freezing obstacles in an irregular fashion, an explicit analytical expression of $\eta$ up to $f_c$ beyond which the Fick-Jacobs description is not valid cannot be obtained. 
However, by looking at the behavior of $\eta$ versus $f$ in 
Figure~\ref{fig:FJ_Validation}, an empirical relation between $\eta$ and $f$ can be obtained as $\eta(f) = c_1 - c_2 \, \erf((f - \lambda)/\Sigma)$. 
After fitting this function to the numerical data, the constant parameters values can be read as 
$c_1 = 0.51, c_2 = 0.41, \lambda = 9$, and $\Sigma = 5$. 
Thus, the rate at which $\eta$ decays exponentially with $f$ is $1/\Sigma$.


\section{effective diffusion}\label{Diffusion}

The effective diffusion is characterized by the mean-squared deviation (variance) of the particle position $x(t)$, i.e., 
$\langle \Delta x^2 (t) \rangle 
= \langle x(t)^2 \rangle - \langle x(t) \rangle^2 $. 
This variance obeys a power law $\sim t^\alpha$, where the power $\alpha$ decides the nature of the particle diffusion. 
If the long time behavior assumes a linear function of time, i.e., $\alpha = 1$, it is a  normal diffusion \cite{Georges_Review}. 
On the other hand, any deviation from the strict linear behavior at asymptotic times is termed as anomalous diffusion \cite{Georges_Review, Klafter_Review}. 
For example, if $0 < \alpha < 1$ it is called as subdiffusion while for $1 < \alpha < 2$ it is called as superdiffusion. 
Note that $\alpha = 0$ corresponds to zero diffusion, i.e., the trapped state.

In the $\eta = 0$ limit, the effective diffusion coefficient can be obtained 
using the Fick-Jacobs equation \cite{Burada_prl, Burada_phd}. 
It is given by
\begin{equation}\label{dif}
\frac{D_{eff}}{D_0}  = \frac{\displaystyle \int_{0}^{1} \int_{x-1}^{x} \frac{D(z)}{D(x)}\, \frac{e^{A(x)}}{e^{A(z)}} \, [I(z)]^2   \, dx \, dz}{\displaystyle \left[\int_{0}^{1}  I(z) \, dz \right]^3}\,,
\end{equation}
where $I(z)$ is same as mentioned before in equation~(\ref{mob}). 
Numerically, for any arbitrary obstacles density, the local exponent and the corresponding local effective diffusion coefficient are, respectively, calculated as \cite{Zeitz}  
\begin{align}
\alpha(t) &:=  \frac{\mathrm{d} \log ( \langle x(t)^2 \rangle - \langle x(t) \rangle^2) }{\mathrm{d} \log t},\label{alpha}\\
D_{eff}(t) &:= \frac{1}{2} \, \frac{\mathrm{d}( \langle x(t)^2 \rangle - \langle x(t) \rangle^2) }{\mathrm{d} t}. 
\label{diffusion}
\end{align}
Note that the steady state values of the exponent and the corresponding effective diffusion coefficient can be obtained as $\alpha = \lim\limits_{t\to\infty} \alpha (t)$ and $D_{eff} = \lim\limits_{t\to\infty} D_{eff}(t)$, respectively.
\begin{figure}[t]
\includegraphics[width=0.5\linewidth]{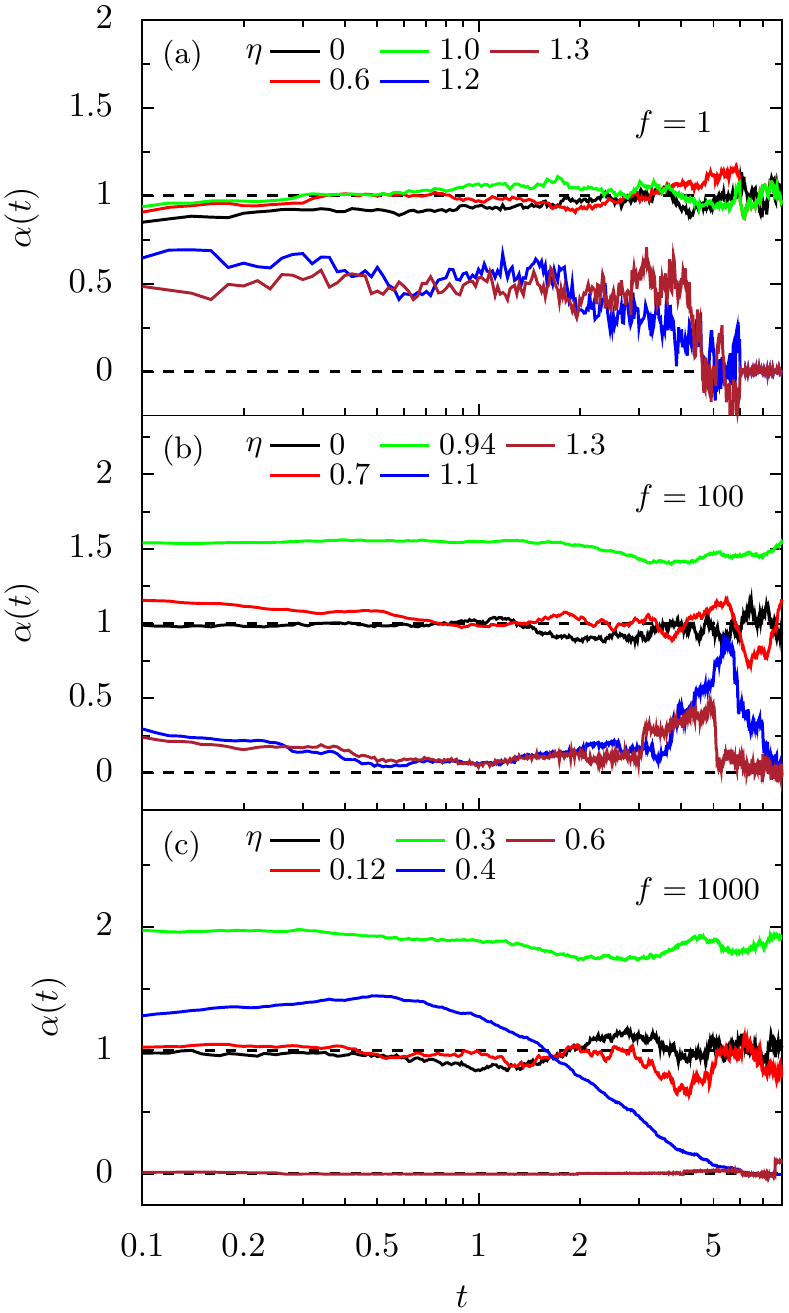}
\caption{(Color online) Local exponent $\alpha(t)$ for various values of the obstacles density $\eta$ and the scaling parameter $f$. 
The other set parameters are $a = 1/2\pi$, $b = 1.02/2\pi$, and $R_0 = 0.02$.}
\label{fig:exponent}
\end{figure} 

Results for the local exponent $\alpha(t)$ are depicted in 
Figure~\ref{fig:exponent} for various values of the obstacles density $\eta$ and  the scaling parameter $f$. 
Note that for the small $f$ value, when $\eta < \eta_c \approx 1.2$, the local exponent fluctuates around one in the long time limit. 
In this limit, as discussed earlier, the thermal fluctuations dominate over the external force, and the particles exhibit normal diffusive behavior.
Whereas, when $\eta \ge \eta_c$, particles get trapped in the percolating clusters, long chains with length in the order of periodicity of the channel. 
As a result, the local exponent becomes zero in the steady state limit. 

On the other hand, for the moderate $f$ value, the obstacles do not influence the normal diffusion of the particles up to $\eta < 0.94$, thus the local exponent fluctuates around unity in the long time limit. 
Interestingly, by further increasing $\eta$, the diffusing particles are trapped partially 
in cavities formed by the obstacles and escape from them due to thermal fluctuations. 
As a result, particles exhibit long jumps, i.e., so called L\'evy flights \cite{West}. 
Correspondingly, the particles show superdiffusive behavior. 
For $\eta \ge 1.1$, as mentioned earlier, the particles get trapped completely in the percolating clusters. Therefore, the local exponent becomes zero in the long time limit.

At the higher value of $f$, for $\eta < 0.12$, as expected, the local exponent fluctuates around one in the long time limit, illustrating that the particles exhibit normal diffusion. 
An increase in $\eta$ leads to partial trapping of particles at the obstacles boundaries for long times because the external force dominates over the thermal fluctuations present in the system which tries to drag the particle along its direction, i.e., a straight line motion.  
As a result, the particles exhibit subdiffusive behavior.  
By further increasing $\eta$, particles are trapped partially in cavities formed by the obstacles and while escaping, they exhibit long jumps which result into 
the superdiffusive behavior. 
As mentioned earlier, the particles get trapped when $\eta \ge 0.4$. Correspondingly, the local exponent becomes zero in the long time limit.
It is interesting to observe that superdiffusive behavior appears just before the trapped state.  
Finally, it is worth to point out that the biased particles exhibit different kinds of diffusive behavior, i.e., normal diffusion, anomalous diffusion, and trapped state, because of the presence of a random Lorentz gas environment in the channel.

\begin{figure}[t]
\includegraphics[width=0.6\linewidth]{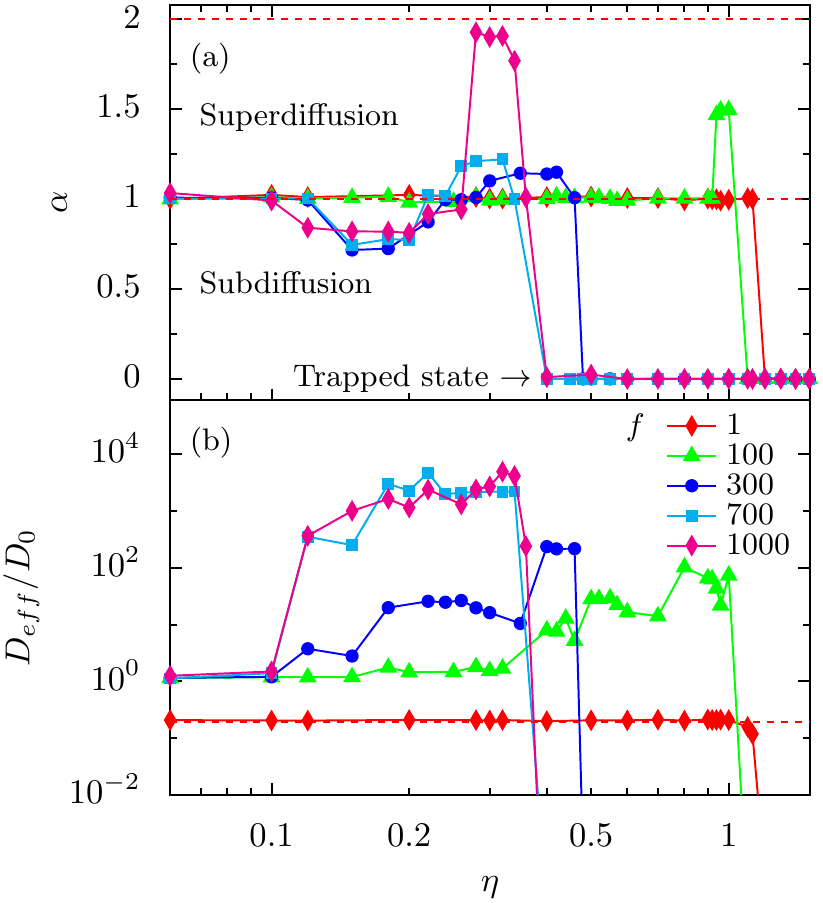}
\caption{(Color online) 
The steady state behaviors of the exponent $\alpha$ (a) and the corresponding scaled effective diffusion coefficient $D_{eff}/D_0$ (b) as a function of the obstacles density $\eta$ for various values of the scaling parameter $f$. 
The dashed line in (b) corresponds to analytical finding for $\eta = 0$ and $f = 1$. 
The other set parameters are $a = 1/2\pi$, $b = 1.02/2\pi$, and $R_0 = 0.02$.}
\label{fig:Phase}
\end{figure} 

Figure~\ref{fig:Phase} depicts the steady state behaviors of the exponent $\alpha$ and the corresponding scaled effective diffusion coefficient $D_{eff}/D_0$ as a function of the obstacles density $\eta$ for various values of the scaling parameter $f$. 
For small $f$, as discussed before, the particles exhibit normal diffusion for $\eta < \eta_c$, and by further increasing $\eta$, the particles get trapped instantaneously (see Fig.~\ref{fig:Phase}(a)). 
Interestingly, the corresponding $D_{eff}/D_0$ remains unaltered and agrees well with the analytical findings (for $\eta = 0$, dashed line) up to $\eta \approx 1$ (see Fig.~\ref{fig:Phase}(b)), reflecting the fact that for this parameter range when the thermal fluctuations dominate over the external force, the obstacles do not have any influence on the effective diffusion of the particles. 
Whereas, $D_{eff}/D_0$ drops to zero for $\eta \ge \eta_c$, indicating that the particles are trapped in the given environment. Note that new features start to emerge as the scaling parameter is increased to moderate and higher values. For the moderate $f$, as explained earlier, the normal diffusion is followed by the superdiffusion, and then the particles get trapped in the given environment well before $\eta_c$. Whereas, for the higher $f$ values, i.e., when $f \ge 300$, the normal diffusion is followed by a transition from subdiffusion to the superdiffusion, and then the particles get trapped in the given environment much more prior to $\eta_c$. 
In particular, for some intermediate $\eta$ values, particles exhibit a normal diffusive behavior during the cross over between either subdiffusion to superdiffusion or superdiffusion to trapped state. 
Here, it is important to point out that the obstacles form percolating clusters inside the channel at the higher obstacles density $\eta$, which slow down the motion of particles for the moderate scaling parameter, i.e., $f = 100$, even though the process is still normal. Therefore, the average survival time of the particles in a cell increases. This behavior plays an important role in the diffusion process, which decreases the nonlinear mobility (see Fig.~\ref{fig:Mobility}) and results in an enhancement of the effective diffusion coefficient (see Fig.~\ref{fig:Phase}(b)). This fact has already been reported in references \cite{Borromeo, Khatri}.
Interestingly, the effective diffusion coefficient is greatly enhanced when the particles exhibit subdiffusive or superdiffusive behavior for various values of $\eta$ at the higher scaling parameter values. 
This is because of the partial trapping of particles for long times at the obstacles boundaries or inside the cavities formed by the freezing obstacles.
However, as expected, $D_{eff}/D_0$ drops down to zero when the particles are trapped.

\section{Conclusions}\label{Conclusions}
  
In this work, we have studied the diffusive behavior of the biased Brownian particles in a two dimensional confined and heterogeneous environment of freezing obstacles modeled by a random Lorentz gas. We have shown that the transport properties and the dynamics of the particles exhibit peculiar features for various values of the obstacles density $\eta$ and the scaling parameter $f$. 
Using Brownian dynamics simulations, we have observed that the particles get trapped or localized in the given environment for any value of the scaling parameter when the system percolates at the critical obstacles density $\eta_c \approx 1.2$, hence for $\eta \ge \eta_c$, the nonlinear mobility and the scaled effective diffusion coefficient become zero. 
Further, we have numerically demonstrated that the particles can be easily trapped prior to the critical obstacles density $\eta_c$ for the moderate and higher scaling parameter values. Interestingly, the nonlinear mobility exhibits a nonmonotonic behavior as a function of the scaling parameter for various values of the obstacles density. In addition, the effective diffusion coefficient shows a greatly enhanced behavior for various values of the obstacles density at the moderate and higher scaling parameter values.  
Moreover, we have obtained a relation between $\eta$ and $f$ which provides an estimate of the minimum $\eta$ up to a critical scaling parameter $f_c$ beyond which the Fick-Jacobs description is invalid.
Also, it has been observed that the particles exhibit normal diffusion, anomalous diffusion, and trapped state, which are genuine to the confined and random Lorentz gas environment, depending on the scaling parameter for various values of the obstacles density.
 
The approach of effectively controlled diffusion of particles in a confined and random Lorentz gas environment by varying the obstacles density could be applied to a wide range of applications, including biochemical reactions in living cells which occur in a complex heterogeneous media \cite{Dix}, 
particles transport in crowded cellular environments \cite{Noe}, 
dispersive transport in disordered semiconductors \cite{Schwarz}, 
controlled drug release \cite{Siegel}, etc.
In the future, the current study can be easily extended in different directions. Important examples we mention, (i) the generalization of freezing obstacles to mobile obstacles with kinetic rate $\kappa$ \cite{Franosch_prl, Basu_Phys,Benichou, Illien_prl} and (ii) the channel walls can be considered as fluctuating walls  \cite{Li_prr}.

\section{Acknowledgements}
 
This work was supported by the Indian Institute of Technology Kharagpur under the Grant No. IIT/SRIC/PHY/TAB/2015-16/114.

\end{document}